\documentclass[12pt]{article}

\usepackage{newtxtext,newtxmath}

\usepackage{graphicx}

\usepackage[letterpaper,margin=1in]{geometry}

\renewenvironment{abstract}
	{\quotation}
	{\endquotation}

\date{}

\makeatletter
\renewcommand{\fnum@figure}{\textbf{Figure \thefigure}}
\renewcommand{\fnum@table}{\textbf{Table \thetable}}
\makeatother

\usepackage{scicite}

\usepackage{url}

\def\scititle{
	Avalanches can increase stored energy in a granular fault
}
\title{\bfseries \boldmath \scititle}

\author{
	A. Douin$^{1}$,
	E. Saurety$^{1}$,
    V. Levy dit Vehel$^{1}$,
    L. Combe$^{1}$,
    L. Vanel$^{1}$, \\
    O. Cochet-Escartin$^{1}$,
		O. Ramos$^{1\ast}$\\
	\small$^{1}$Universit\'e Lyon 1-CNRS, Institut Lumi\`ere Mati\`ere, UMR5306\\
	\small 69622 Villeurbanne, France.\\
	\small$^\ast$Corresponding author. Email: osvanny.ramos@univ-lyon1.fr
}

\begin{document} 

\maketitle

\begin{abstract} \bfseries \boldmath
Slowly sheared granular materials generally store mechanical energy and dilate between abrupt failures that release energy and compact the material. Here, simultaneous measurements of torque, layer thickness, acoustic emission, and photoelastic force networks reveal all four combinations of energy release or storage with contraction or dilation in a compressed granular fault. Most strikingly, some avalanches both dilate the layer and increase the elastic energy transmitted to the resisting boundary. These events reorganize force chains beyond the shear band and produce a distinct acoustic response. The results show that an avalanche need not relax a driven disordered material; it can instead redistribute stress into a more highly loaded configuration, a mechanism with potential relevance to the physics of both laboratory and natural faults.
\end{abstract}

\noindent
First described by Reynolds in 1885, slowly shearing a granular material under constant pressure --what we now recognize as a granular fault-- causes it to expand, a phenomenon known as dilatancy \cite{Reynolds1885}. As shear proceeds, the system evolves through a succession of mechanically stable configurations, progressively storing mechanical energy and dilating, until it suddenly becomes unstable and collapses \cite{Kolb2004, Lherminier2019}. This abrupt event reorganizes the grains into a new stable state. Over the past 25 years, such jammed states have been widely investigated \cite{Liu1998, Kolb2004, Dauchot2005, Majmudar2007, Behringer2019}, driven by analogies with the glass transition \cite{DAnna2001, Mari2009, Parisi2010, Charbonneau2017} and by major advances in idealized, frictionless models \cite{Parisi2010, Behringer2019}. Real granular materials, however, are inherently frictional, leading to history-dependent, out-of-equilibrium behavior that limits the applicability of traditional statistical mechanics \cite{Parisi2010}.

Sudden structural reorganizations, commonly termed avalanches, present an even greater challenge. Initiated at grain–grain contacts, cascades of frictional failures \cite{BenDavid2010} and collisions can propagate throughout the system. Granular avalanches have therefore attracted considerable attention, both as a fundamental problem in granular physics \cite{Daerr1999, Jop2006} and as laboratory analogs of earthquakes \cite{Johnson2005, Daniels2008, Riviere2018, Zadeh2019b, Lherminier2019}. Under shear, they typically occur either as relatively regular events with a characteristic size \cite{Johnson2005}, or as scale-free events \cite{Daniels2008, Zadeh2019b, Lherminier2019} spanning many orders of magnitude in energy \cite{Lherminier2019}. While modern machine-learning approaches can successfully anticipate large events in the former case \cite{RouetLeduc2017}, prediction remains elusive for scale-invariant avalanches.

Because a granular fault gradually stores energy and dilates under quasistatic shear, avalanches are generally expected to relax the system and induce contraction. Whether a rapid internal reorganization can instead increase the energy stored by the fault remains unresolved. Here, simultaneous boundary, acoustic, and grain-scale measurements show that avalanches can combine dilation with increased elastic loading, challenging the conventional picture of failure as a purely relaxing process.

\subsection*{Experimental Setup}

We constructed a granular fault consisting of a single layer of approximately 4,000 photoelastic disks (4 mm thick), 3D-printed in two different diameters and confined within a 5 mm gap between two concentric, fixed cylindrical walls. The disks are vertically confined by two mobile annular plates patterned with half-disks. The lower plate rotates at a constant speed of 48 mm/h, while the upper plate is prevented from rotating but is free to move vertically, compressing the granular layer under a dead load of approximately 20 kg.

Under slow shear and compression, mechanical energy accumulates gradually and is intermittently released through sudden structural reorganizations spanning a wide range of sizes \cite{Lherminier2019} . These events, associated with acoustic emissions, exhibit earthquake-like behavior \cite{Duplat2025}.
A shear band is formed at a distance of about ten grain diameters from the rotating plate \cite{Kuwano2013, Henann2013, Shekari2023}. 

We continuously record the force required to drive the lower plate ($F_{\mathrm{drive}}$), the shear force exerted by the grains on the upper plate ($F_{grains}$), and the vertical displacement of the upper plate ($h$) at 10 kHz. In parallel, six piezoelectric sensors record acoustic emissions at 100 kHz, while 24 synchronized cameras image the structure every 4 s. Forces are converted into torques, $\Gamma_{\mathrm{drive}}$ and $\Gamma_{grains}$, considering the radial distance of the disks from the cylinder center; and all mechanical and acoustic measurements are expressed in terms of energy:  $E_{\mathrm{drive}}$, $E_{grains}$, $E_{h}$ and $E_{ac}$. Each experimental run lasts approximately 100 hours and captures several hundred thousand avalanches.

\subsection*{Global dynamics of the granular fault}

\begin{figure}
	\centering
	\includegraphics[width=6.5 in]{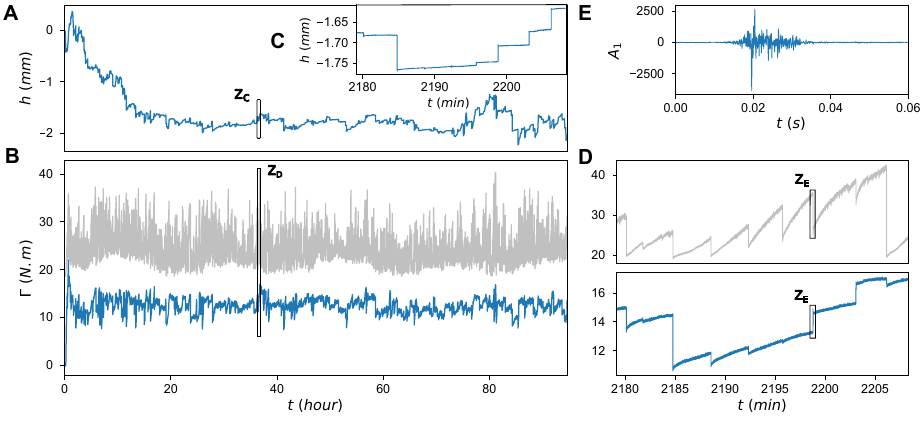}  
	\caption{\textbf{Mechanical and acoustic signals measured during a 4-day experiment}
		(\textbf{A}) Vertical position $h(t)$ of the upper plate over the four-day reference experiment, showing intermittent dilation and contraction episodes associated with granular rearrangements. Inset (\textbf{C}) : close-up view around a major events, highlighting the abrupt displacement characteristic of large avalanches. (\textbf{B}) Time series of the driving-chain torque (gray) and resisting torque (blue) recorded during the same experiment. (\textbf{D}): zoom on events' time series of the driving-chain torque (gray) and resisting torque (blue), showing the slow increase interrupted by rapid drop in mechanical energy and stress release. The time x-axis is shared for both plots. The y-axis label is shared for (\textbf{B}) and (\textbf{C}) plots. (\textbf{E}) Acoustic signal from sensor 1 during a representative main event ($Z_D$), revealing a burst of high-frequency activity coincident with the mechanical instability. The mechanical response of the sheared granular layer is quantified through simultaneous measurements of torque at both end of the fault and fault dilation.}
	\label{fig:Raw_datas} 
\end{figure}

$\Gamma_{grains}$ exhibits a transient of approximately five hours before reaching a steady state (Fig.~\ref{fig:Raw_datas}.B). In contrast, the evolution of dilatancy $h$ is significantly slower (Fig.~\ref{fig:Raw_datas}.A): the system first undergoes dilation, associated with an increase in $\Gamma_{grains}$, followed by a slow global contraction, before reaching a steady state about 15 hours after the start of the experiment.
The global dynamics of both $\Gamma$ signals display stick–slip behavior, characterized by a slow increase in force punctuated by small jumps. Variations in the slope reflect substantial changes in the elastic response of the granular layer, while sudden torque drops (avalanches) correspond to abrupt releases of elastic energy (Fig.~\ref{fig:Raw_datas}.D) accompanied by acoustic bursts (Fig.~\ref{fig:Raw_datas}.E). Each avalanche drives the system into a new jammed state, from which a new loading phase begins. The global dilatancy $h$ shows a similar pattern, with gradual dilation interrupted by abrupt changes associated with avalanches (Fig.~\ref{fig:Raw_datas}.C).

Surprisingly, not all avalanches are associated with torque drops or contraction; many of the biggest events instead exhibit sudden $\Gamma_{grains}$ increases (Fig.~\ref{fig:Raw_datas}.D) or abrupt dilation, potentially leading to an abrupt increase in the energy stored within the fault. 

\subsection*{Avalanche modes}

Across the full dataset, we identify approximately $4\times10^{5}$ acoustic bursts, $10^{5}$ mechanical avalanches, and $2\times10^{4}$ discrete displacement events. We characterize each avalanche by the associated variations in mechanical energy at the driving and resisting boundaries, $\Delta E_{\mathrm{drive}}$ and $\Delta E_{\mathrm{grains}}$, respectively, and by the normal contribution associated with the upper-plate displacement, $\Delta E_{\mathrm{h}}$. Because the driving boundary is the imposed control parameter, $\Delta E_{\mathrm{drive}}$ is systematically negative. In contrast, $\Delta E_{\mathrm{grains}}$ and $\Delta E_{\mathrm{h}}$ can have either sign. A negative $\Delta E_{\mathrm{grains}}$ corresponds to elastic-energy release by the granular fault, whereas a positive value indicates an abrupt increase in the force transmitted to the resisting sensor, and hence an increase in stored mechanical energy. Similarly, $\Delta E_{\mathrm{h}}<0$ and $\Delta E_{\mathrm{h}}>0$ correspond to contraction and dilation of the granular layer, respectively.

The resisting torque $\Gamma_{\mathrm{grains}}$ and the upper-plate displacement $h$ probe, respectively, the radial and vertical projections of the force-chain network acting on the upper boundary. The mechanical energy variation at the upper plate can therefore be decomposed as

\begin{equation}
\Delta E = \Delta E_{\mathrm{grains}}+\Delta E_{\mathrm{h}},
\end{equation}

where the two terms describe the tangential and normal contributions to the energy exchanged during an avalanche. Energy variations measured at both boundaries exhibit scale-free distributions (Fig.~\ref{fig:sup_events_characterization}.A).

\begin{figure}
	\centering
	\includegraphics[width=6.5 in]{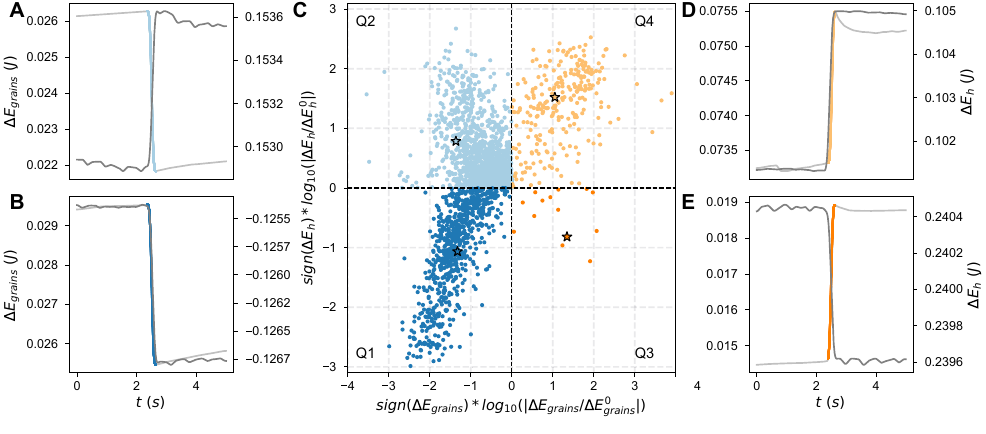}  
	\caption{\textbf{Four coupled modes of avalanches behavior.}
		(\textbf{C}) Relation between the signed and normalized vertical potential energy evaluated with the displacement $\Delta E_{\mathrm{h}}$ and the signed and normalized potential energy evaluated from the resisting torque variation $\Delta E_{\mathrm{grains}}$, for all matched events, showing the four dynamical regimes emerging from this joint response. Colors correspond to the quadrant category : $Q1$ (blue), $Q2$ (light blue), $Q3$ (orange), $Q4$ (light orange). The star marker correspond to the selected event to display in (\textbf{A}) to (\textbf{E}).  (\textbf{A-B-D-E}) Zooms on the vertical position of the upper plate (gray) and on the resisting torque mechanical energy (silver) for representative events belonging to the four dynamical quadrants. The time x-axis, the torque y-axis and the position twin y-axis are shared for all plots except (\textbf{C}).}
	\label{fig:Quadrants} 
\end{figure}

We classify avalanches according to the joint signs of $\Delta E_{\mathrm{grains}}$ and $\Delta E_{\mathrm{h}}$, defining four distinct modes (Fig.~\ref{fig:Quadrants}.B). $Q1$ corresponds to the intuitive dissipative-collapse regime, with $\Delta E_{\mathrm{grains}}<0$ and $\Delta E_{\mathrm{h}}<0$: the system releases stored elastic energy while contracting. $Q2$ and $Q3$ correspond to mixed responses, with $(\Delta E_{\mathrm{grains}}<0,\Delta E_{\mathrm{h}}>0)$ and $(\Delta E_{\mathrm{grains}}>0,\Delta E_{\mathrm{h}}<0)$, respectively. In $Q4$, both quantities are positive, corresponding to the counterintuitive regime in which the granular layer dilates while the mechanical energy stored at the resisting boundary increases.

We refer to events with $\Delta E_{\mathrm{grains}}>0$ as ``negative avalanches'' because their inferred energy release is negative. In addition to these coupled modes, individual observables can exhibit pure responses, corresponding to isolated compaction or dilation in $h$, or to isolated drops or increases in the resisting mechanical energy. Although coupled events are less frequent, on average they carry the largest energy variations (Fig.~\ref{fig:sup_events_characterization}.B).

Pure energy drops constitute the most frequent events, consistent with the expected dissipative response of a compacted, frictional granular layer. Along the same lines, the second most frequent class is $Q2$, in which dilation is accompanied by a decrease in stored mechanical energy. The largest avalanches are predominantly found in $Q1$, corresponding to the conventional picture of a collapse accompanied by elastic-energy release, but substantial high-energy events also occur in $Q4$, where dilation coincides with energy accumulation. The prevalence of such negative-energy-release events challenges the conventional picture of granular avalanches as purely dissipative failures and points instead to a coupled interplay between dilatancy, force-chain reorganization, and stress storage. 

\subsection*{Impact of Events on Fault Behavior} 

To quantify the relationship between the mechanical response and acoustic emission, we perform linear regressions in log-log space between the acoustic energy $E_a$ and the mechanical energy variations $\Delta E_{\mathrm{grains}}$ and $\Delta E_{\mathrm{h}}$. The acoustic energy exhibits little dependence on the sign of the vertical displacement $\Delta h$, with only a weak difference in amplitude: $Q4$ events emit slightly more acoustic energy on average. In contrast, $E_a$ depends strongly on the sign of $\Delta E_{\mathrm{grains}}$ for the two categories considered. Events associated with an increase in resisting torque ($\Delta E_{\mathrm{grains}}>0$, corresponding predominantly to $Q4$) systematically depart from the global scaling relation. Their acoustic energy exhibits a weaker dependence on the torque variation and remains systematically higher than the global trend at small $\Delta E_{\mathrm{grains}}$.

To quantify the temporal impact of a large event, we analyze the mechanical response surrounding the 150 largest events in each quadrant. For each quadrant, we compare the time-resolved averages of $\Delta E_{\mathrm{drive}}$, $\Delta E_{\mathrm{grains}}$, $\Delta\Gamma_{\mathrm{drive}}^{\ 2}$, and the loading rate $k_{\mathrm{drive}}$.

\begin{figure}
	\centering
	\includegraphics[width=6.5 in]{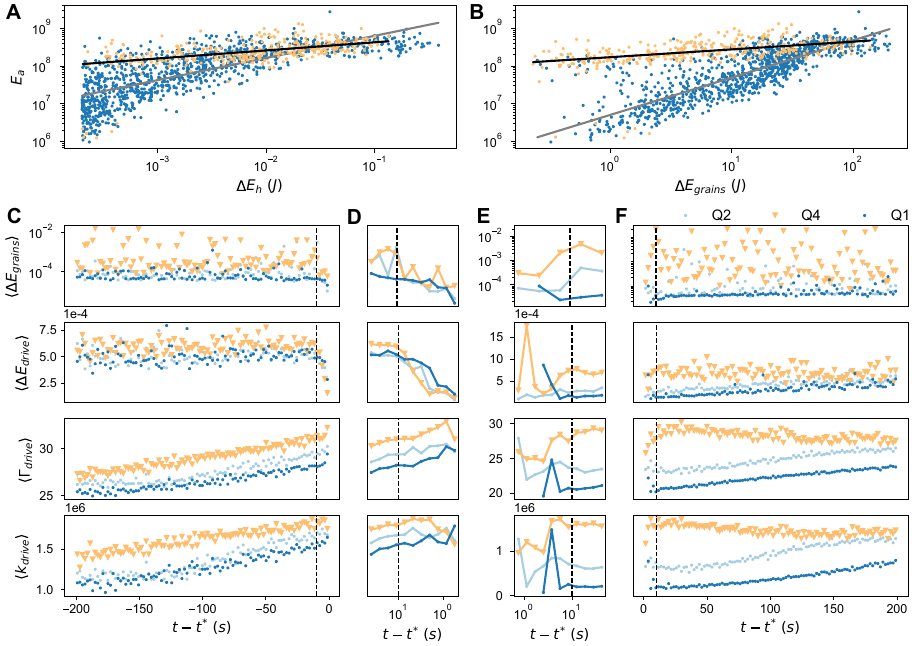} 
	\caption{\textbf{Impact of avalanches' mode on fault dynamics}
		(\textbf{A-B}) Scatter plot of the emitted acoustic energy $E_{\mathrm{a}}$ per event with respect to the potential energy variation $\Delta E_{\mathrm{h}}$ and the mechanical energy variation $\Delta E_{\mathrm{grains}}$, for $Q1$ (blue) and $Q4$ (light orange) events. A log power-law fit is done on the $Q1$ (gray) and $Q4$ (black) data. Y-axis is shared by both plots. (\textbf{C-F}) Average over 150 events per category of : energy variation at both en of the fault $\Delta E_{\mathrm{drive}}$ and $\Delta E_{\mathrm{grains}}$, driving torque $\Gamma_{\mathrm{drive}}$ and the shear band loading rate $k_{\mathrm{drive}}$, with respect to the time delay to event $t - t*$. $t*$ mark the start of the event and $t - t*  = \pm 200$  correspond roughly to the auto-correlation time of the mechanical response. Black dash line mark $t - t* = 10$ seconds. Y-axis are share horizontally, and x-axis are shared vertically. (\textbf{D-E}) Zoom right before (resp. after) main event at $t*$.}
	\label{fig:Fault_dynamic} 
\end{figure}

Events surrounding $Q4$ main shocks are systematically larger and occur under higher torque levels and loading rates than those associated with $Q1$ and $Q2$ events (Fig.~\ref{fig:Fault_dynamic}.A and~\ref{fig:Fault_dynamic}.B). These differences progressively fade away with temporal distance from the main shock. Regardless of the quadrant, the immediate pre-failure regime is marked by a pronounced reduction in precursor activity and a decrease in loading slope. After the main event, correlations decay progressively, while the short-time peaks originate from oscillatory responses of the apparatus.

\subsection*{Structural Response}

To identify the structural signatures of the four avalanche regimes, we reconstruct the grain contact network from panoramic images and track grain displacements across individual events (\cite{methods}). We analyze the 150 largest events in each quadrant: $Q1$--$Q4$. Energy-releasing events ($Q1$ and $Q2$) are characterized by pronounced grain motion concentrated within the shear band, with coherent vortical structures. In the jammed region they exhibit a net displacement in the radial direction consistent with local compaction, alongside a slight backward displacement (Fig.~\ref{fig:Structure}.A and Fig.~\ref{fig:Structure}.B; Fig.~\ref{fig:sup_img_analysis}.D). By contrast, energy-storing events ($Q3$ and $Q4$) produce displacements in the shear band that are nearly two times smaller and lack a well-defined deformation pattern. Their structural signature instead extends into the otherwise jammed region, where grain motion does not reverse direction along the shear axis and increasing $\Delta y$ is consistent with local dilation. 

The force network provides a complementary view of this distinction (Fig.~\ref{fig:Structure}.C). For $Q1$ and $Q2$, the mean variation of the photoelastic contact forces, $\langle\Delta F_{\mathrm{photoelastic}}\rangle$, is negative, consistent with a net release of stored force. For $Q3$ and $Q4$, the mean variation is close to zero, indicating little net force release. Yet their mean absolute variation, $\langle|\Delta F_{\mathrm{photoelastic}}|\rangle$, remains substantial, particularly for $Q4$, revealing pronounced local rearrangements of the force network despite the comparatively weak grain displacements. 

\subsection*{Interpretation}

Mechanical energy is continuously injected into the system by the slow rotation of the driving boundary. This loading progressively deforms the granular layer and increases the stresses stored within its contact and force-chain network. Superimposed on this quasi-affine response, reflected by the gradual increase of the resisting torque, abrupt fluctuations arise as the granular structure undergoes rapid reorganizations. These fluctuations define the avalanches and provide direct access to the evolution of the internal force-bearing skeleton.

The origin of this rich dynamics lies in the extended network of force chains percolating across the two-dimensional granular fault (Fig.~\ref{fig:sup_img_analysis}), which collectively supports the imposed normal and shear stresses. The photoelastic measurements show that avalanches are not solely associated with the weakening of this network, but, more importantly, with its reorganization. Some events lead to a decrease in resisting torque and release elastic energy stored in the granular assembly, whereas others produce an increase in torque and therefore an apparent accumulation of elastic energy. These ``negative avalanches'' do not result from a sudden injection of energy: the motor imposes a constant loading rate and therefore does not abruptly increase the stress applied at the driving boundary. Instead, the event changes how the applied stresses are transmitted through the granular packing, modifying the configuration and loading of the force-bearing network.

\begin{figure}
	\centering
	\includegraphics[width=6.49 in]{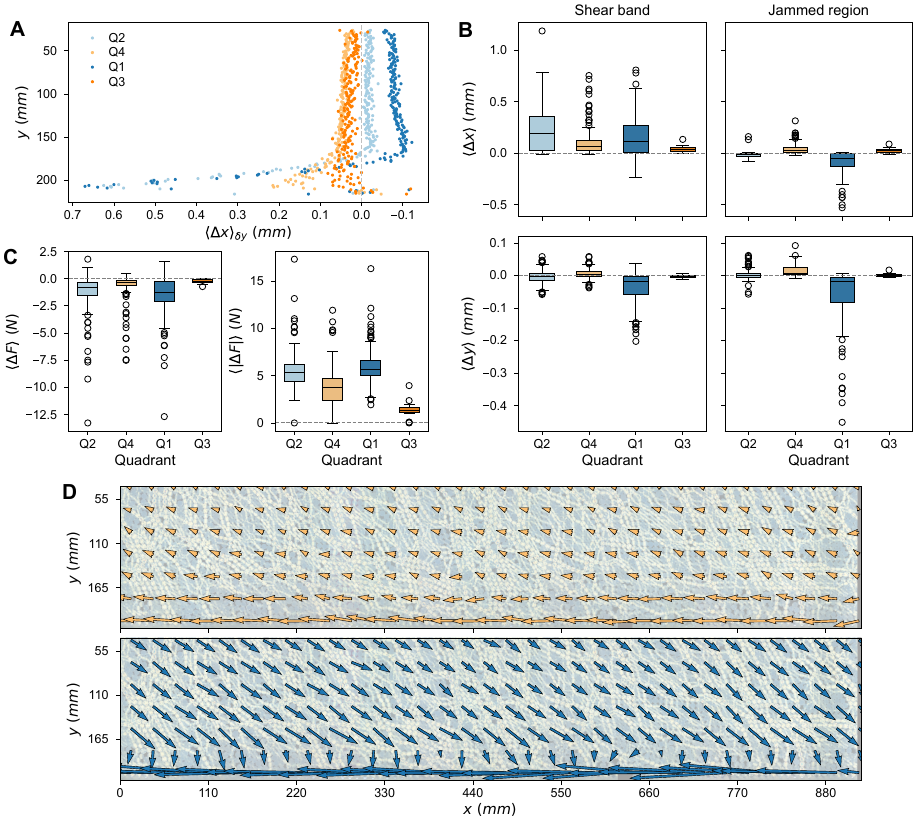} 
	\caption{\textbf{Structural Response}
		\textbf{(A)} Mean grain displacement along the driving direction, $\langle \Delta x(y) \rangle$, as a function of the vertical position $y$ within the granular layer. Displacements are averaged over vertical bins of thickness $\delta y$ and over the 150 largest events in each category, $Q1$--$Q4$. The gray dashed line denotes zero displacement. (\textbf{B}) Box plot of the mean grain displacement per event, along the driving direction ($\Delta x $) and the transverse one ($ \Delta y $), for both the shear band and the jammed region, with respect to the event category. The gray dashed line denotes zero displacement. Y-axis are share horizontally, and x-axis are shared vertically. (\textbf{C}) Box plot of the average contact's force variation around an event for each category and its absolute value, averaged on all contacts. (\textbf{D}) Coarse grained displacement filed average on 150 events for $Q4$ (light orange) and $Q1$ (blue), display on top of reference panorama image. X-axis are shared by both fields.}
	\label{fig:Structure} 
\end{figure}

The coupling between force-chain reorganization, dilatancy, and mechanical energy gives rise to the four avalanche classes identified above. Energy-releasing events ($Q1$ and $Q2$) are associated with a substantial weakening of the force network and a release of stored elastic energy. Their structural response is concentrated primarily within the shear band, where large grain displacements and coherent vortical structures accommodate the imposed slip (Fig.~\ref{fig:Structure}.A, B and Fig.~\ref{fig:sup_img_analysis}.D). In contrast, if considered only within the shear band, energy-storing events ($Q3$ and $Q4$) exhibit displacement amplitudes that are nearly two times smaller and lack a clearly organized deformation pattern. But their response also extends into the jammed region, with a positive displacement aligned with the shear direction, blurring the distinction between the shear band and the jammed region. Thus, both the magnitude and direction of grain displacement depend on the avalanche category and on the grain position across the layer, revealing a depth-dependent response as well as two distinct modes of structural reorganization associated with energy release and energy storage.

The force network provides a complementary signature of this distinction. For $Q1$ and $Q2$, the mean variation of the photoelastic contact forces, $\langle\Delta F_{\mathrm{photoelastic}}\rangle$, is negative, consistent with a net release of force stored in the granular network. For $Q3$ and $Q4$, the mean variation is close to zero, indicating little net force release. Yet their mean absolute variation, $\langle|\Delta F_{\mathrm{photoelastic}}|\rangle$, remains substantial, particularly for $Q4$, revealing pronounced local rearrangements of the force network despite the comparatively weak grain displacements. Together, these observations show that avalanches can reorganize the granular state through fundamentally different combinations of particle motion and force-network rearrangement. These observations suggest that the two classes arise from distinct reorganizations at the boundary of the shear band. For $Q1$ and $Q2$, the force network locally loses its ability to sustain the applied stress, allowing the shear band to slip while the surrounding jammed region responds in the opposite direction as stored elastic energy is released. In $Q1$, this relaxation is accompanied by local compaction, whereas in $Q2$ it occurs together with dilation. Conversely, for $Q3$ and $Q4$, the local force-chain topology appears to prevent such a direct rupture. The region involved in the rearrangement therefore extends beyond the pre-existing shear band and progressively incorporates the jammed region into the deformation. In $Q4$, this collective reorganization is accompanied by dilation and a pronounced rearrangement of the force network, while the much rarer $Q3$ events retain a response compatible with local compaction. The latter are predominantly observed early in the experiment, when sufficient free volume remains within the granular layer; as the system evolves, this free volume progressively decreases, making this mode of compaction increasingly difficult.

The acoustic response provides an independent signature of this asymmetry. Although acoustic and mechanical energies exhibit consistent scaling at the population level, their amplitudes differ systematically between events with opposite signs of torque variation, whereas no comparable distinction is observed with the sign of the displacement variation. The dominant distinction is therefore not dilation versus contraction, but whether the event releases or stores mechanical energy. This asymmetry can be understood from the spatial distribution of the rearrangements. In events dominated by the shear band, such as $Q1$ and $Q2$, much of the structural reorganization occurs relatively far from the acoustic sensors. In addition, the rupture of force chains along which elastic disturbances propagate may attenuate the measured acoustic signal \cite{Lherminier2014}. In $Q4$, by contrast, the rearrangement extends into the jammed region, closer to the sensors, while the force network remains largely connected rather than undergoing extensive rupture. This geometry provides a possible explanation for the enhanced acoustic amplitude observed for energy-storing events.

The temporal evolution around major events provides evidence that the granular fault retains a memory of its loading history. Before an event, the system exhibits a quiescent phase accompanied by a positive curvature of the resisting-torque signal, consistent with the accumulation of subthreshold rearrangements and the presence of weak precursors. Importantly, however, we find no systematic signature preceding one avalanche class over another, furthermore when fluctuations around the mean signal are taken into account. The distinction between avalanche classes instead emerges primarily from the state left behind by the event. In particular, $Q4$ events combine little net force release with substantial force-network reorganization, dilation, and an increase in stored mechanical energy. They therefore leave the fault in a more strongly loaded state, potentially closer to a subsequent instability. This post-event state may explain the enhanced tendency of $Q4$ events to be followed by further failure and suggests that avalanche predictability is governed less by class-specific precursors than by the state of the fault after an event.

More broadly, our observations suggest that whether the shear band can slip independently of the surrounding jammed region is controlled not only by the boundary conditions, but also by the local topology of the contact network. Energy-releasing events appear to require a configuration that permits the force network to rupture at the shear-band boundary, allowing the shear band to decouple from the jammed region. When locally ordered configurations instead stabilize the contact network against such rupture, the two regions may remain mechanically coupled and reorganize collectively, causing the effective boundary of the shear band to shift into the surrounding material. This suggests that the position and dynamics of a shear band need not be fixed properties of the material, but can emerge dynamically from the instantaneous internal configuration of its contact network. Identifying the structural motifs that favor rupture or, conversely, stabilize the network could therefore provide a route to predicting avalanche modes through graph-based analyses of the contact network, including stability analysis or graph neural networks designed to identify fragile and mechanically resistant configurations \cite{berthier_forecasting_2019}. More generally, this mechanism highlights how complex internal force and contact networks can mediate the response of driven disordered systems to boundary loading, and may provide a framework for understanding analogous shifts in failure localization in other experimental systems and, potentially, in natural faults where the underlying stress and force networks are considerably more complex and difficult to access.

The quantitative event frequencies and scaling relations may depend on the confinement geometry, loading protocol, and detection thresholds. Testing these dependencies across independent realizations will be necessary before extending the proposed mechanism to natural faults.


\clearpage 

%
\bibliography{science_template} 
\bibliographystyle{sciencemag}

%
%
%
%
%
%


\section*{Acknowledgments}
O.R. thanks K. J. M{\aa}l{\o}y and E. Altshuler for their support, which was essential for the building up of the project.  

\paragraph*{Funding:}
This work was funded by the ANR grant ANR-22-CE30-0046 and the CNRS-MITI Rare events: "BigLabQuakes" grant. 

\paragraph*{Author contributions:}
O.R. conceived the project. E.S., V.L., L.C., and O.R. performed the experiments. V.L., L.C., L.V., and O.R. analyzed a preliminary version of the experimental data. A.D. developed the analysis pipeline integrating mechanical, acoustic, and structural information. E.S. developed the image and granular-network analyses. A.D., E.S., O.C., and O.R. analyzed the results and wrote the manuscript.

\paragraph*{Competing interests:}
There are no competing interests to declare.

\paragraph*{Data and materials availability:}

All raw and processed data are available under CC-By licence at \cite{LabQuakes_150} in the LabQuakes Project. 
The source code of the data analysis can be found at \cite{LabQuakes_up}. 

\subsection*{Supplementary materials}
Materials and Methods\\
Figs. S1 to S3


\newpage


\renewcommand{\thefigure}{S\arabic{figure}}
\renewcommand{\thetable}{S\arabic{table}}
\renewcommand{\theequation}{S\arabic{equation}}
\renewcommand{\thepage}{S\arabic{page}}
\setcounter{figure}{0}
\setcounter{table}{0}
\setcounter{equation}{0}
\setcounter{page}{1} 


\begin{center}
\section*{Supplementary Materials for\\ \scititle}

A. Douin$^{1}$,
E. Saurety$^{1}$,
V. Levy dit Vehel$^{1}$,
L. Combe$^{1}$,
L. Vanel$^{1}$,\\
O. Cochet-Escartin$^{1}$, 
O. Ramos$^{1\ast}$ \\ 
\small$^\ast$Corresponding author. Email: osvanny.ramos@univ-lyon1.fr \\
\end{center}

\subsubsection*{This PDF file includes:}
Materials and Methods\\
Figures S1 to S3

\newpage


\subsection*{Materials and Methods}




\subsubsection*{System apparatus}

The bidisperse disks are confined between two transparent, concentric acrylic cylinders (inner diameter~28~cm, outer~29~cm, gap~5~mm). The disks, 3D-printed in photoelastic Durus White 430 (Objet30 printer), have a thickness of~4~mm and diameters of~6.4~mm and~7.0~mm in equal proportion to prevent crystallization. Their high Young’s modulus ($\approx100$~MPa) compared to classical photoelastic materials ($\approx4$~MPa) (4, 12).The granular layer is bounded by two rough, 3D-printed rings composed of 99 half-cylinders (diameter $=6.4$~mm, spacing $\approx\surd2d\approx9$~mm).

The bottom ring, connected through a gear mechanism and roller chain, is driven by a stepper motor whose rotation is reduced by a factor of 2200, resulting in a quasi-static shear with a period of 18.33~hours (linear velocity $48.8$~mm\,h$^{-1}$). The chain tension is maintained by a calibrated spring ($k_{gear}=230$~kN\,m$^{-1}$), defining the stiffness of the apparatus. 

The mechanical response of the granular fault is monitored through simultaneous measurements of the the driving torque (at the bottom plate where control shearing velocity is applied) and the resisting torque and layer dilation at the top plate. Torque and position signals are continuously acquired at 10~kHz using the same data-acquisition system (NI-USB-6366), ensuring sub-millisecond synchronization between stress and volumetric fluctuations. 

Acoustic emissions are monitored by six piezoelectric transducers (VP-1.5, CTS Valpey Corp.) embedded in the upper ring and acoustically coupled to the granular layer using silicone oil. Signals are acquired at 100~kHz on an independent data-acquisition system. 

In parallel, 24 Raspberry-Pi cameras image the granular assembly at 0.5~Hz. Image acquisition is synchronized with both mechanical and acoustic recordings through a common trigger signal, providing a fully synchronized view of stress evolution, grain-scale rearrangements, and energy dissipation during more than three days of continuous shear.

\subsubsection*{Mechanical Response}

The driving torque, $\Gamma_{\mathrm{drive}}(t)$, is measured using a force sensor mounted on the motor assembly and calibrated through the application of known loads. The sensor output voltage is digitized as signed 16-bit integers, with $2^{15}$ counts corresponding to 10~V. Owing to the geometry of the chain transmission, the force applied to the granular layer is related to the motor force by

\begin{equation}
\vec{F}_{\mathrm{drive}}
=
\frac{r}{R}
\frac{\sqrt{r^{2}+D^{2}}}{D}
\vec{F}_{\mathrm{motor}},
\label{eq:motor_to_chain}
\end{equation}

from which the torque transmitted to the fault is directly inferred. The resisting torque exerted by the granular layer, $\Gamma_{\mathrm{grains}}(t)$, is measured using a calibrated steel lever coupled to an Interface SML-900N force sensor (900~N range, stiffness $1.1\times10^{7}$~N\,m$^{-1}$) positioned 55~mm from the center of the upper plate. To compare stress fluctuations at both boundaries of the fault, force measurements are converted into equivalent torques accounting for the transmission geometry. At the upper boundary, the force recorded by the sensor, $\vec{F}_{\mathrm{resistive}}$, is related to the force transmitted by the granular layer through

\begin{equation}
\vec{F}_{\mathrm{grains}}
=
\frac{r}{R}
\vec{F}_{\mathrm{resistive}},
\label{eq:resistive_to_grains}
\end{equation}

where $r$ denotes the lever arm of the force sensor and $R$ the radius of the upper ring.

The dilation of the granular layer is monitored through the vertical displacement of the upper plate, $h(t)$, measured with a capacitive sensor. Sensor calibration is performed before each experiment by zero alignment and displacement referencing using calibrated 0.1~mm spacers. A second sensor independently verifies the verticality of the plate motion, consistent with image-based measurements.

\subsubsection*{Mechanical Energy Evaluation}

Between successive avalanches, the granular fault is loaded quasi-statically at a constant angular velocity $\omega$. During these loading phases, the apparatus behaves as an assembly of coupled elastic elements, including the driving mechanism, the granular layer, and the resisting force sensor. Under the angular spring approximation ( $x^2\sim\left(\omega t\right)^2$), the elastic energy stored in a torsional element of stiffness $k$ is

\begin{equation}
E=\frac{1}{2}k_\theta\theta^2
=\frac{1}{2}r^2k\theta^2
=\frac{\Gamma^2}{2r^2k},
\label{eq:meca_nrj_explain}
\end{equation}

where $k_\theta=r^2k$ is the angular stiffness, $\theta=\omega t$ the angular displacement, and $\Gamma=rF$ the transmitted torque. Between avalanches, we made the first order approximation of a constant loading rate, implying affine increase of the torque:

\begin{equation}
\Gamma=r^2k\omega t,
\end{equation}

which is verified experimentally over several decades in time.

The signed elastic-energy change during an avalanche is therefore obtained directly from the torque discontinuity,

\begin{equation}
\Delta E_m
=
\frac{\Gamma_f^2-\Gamma_i^2}
{2r^2k_{\mathrm{eff}}},
\label{eq:meca_nnrj_def}
\end{equation}

where $\Gamma_i$ and $\Gamma_f$ denote the torque immediately before and after the event, respectively, and $k_{\mathrm{eff}}$ is the effective stiffness evaluated from the preceding loading phase (Supplementary Section~D).

Applying Eq.~(\ref{eq:meca_nnrj_def}) to the two torque measurements yields

\begin{equation}
\Delta E_{\mathrm{drive}}
=
\frac{\Gamma_{\mathrm{drive}}^2(t_f)-\Gamma_{\mathrm{drive}}^2(t_i)}
{2R^2k_{\mathrm{drive}}},
\label{eq:meca_nrj_lower_def}
\end{equation}

for the driving boundary, and

\begin{equation}
\Delta E_{\mathrm{grains}}
=
\frac{\Gamma_{\mathrm{grains}}^2(t_f)-\Gamma_{\mathrm{grains}}^2(t_i)}
{2R^2k_{\mathrm{grains}}},
\label{eq:meca_nrj_upper_def}
\end{equation}

for the resisting boundary.

Accordingly, negative values of $\Delta E$ correspond to elastic-energy release, whereas positive values indicate that the granular structure stores additional elastic energy during the rearrangement. These ``negative avalanches,'' whose inferred energy release is negative, constitute one of the central observations of the present work.

\begin{figure}
	\centering
	\includegraphics[width=6.5 in]{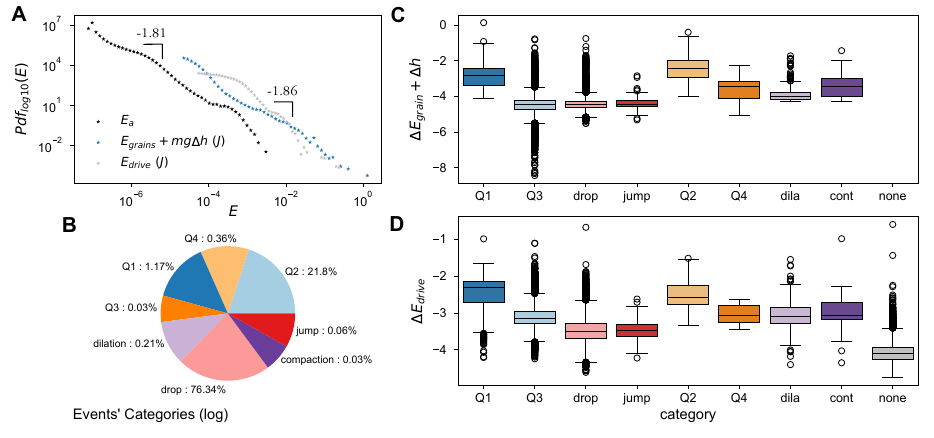} 
	\caption{\textbf{Events' detection}
		(\textbf{A}) Probability density function of events' energy, estimated from: acoustic emission (in black), energy variation at the bottom plate (in gray), and energy variation at the top plate (in blue). All energy variations follow power laws over more than four decades. (\textbf{B}) Events' category proportion on a logarithmic scale with actual percentages. Single-mode behavior represents 76.64\% and coupled-mode behavior 23.36\%. (\textbf{C} and \textbf{D}) Box plot of the logarithm of the absolute energy variation at both ends of the fault with respect to event category. Couple mode behavior includes, on average, the most energetic events.}
	\label{fig:sup_events_characterization} 
\end{figure}

The elastic energies measured at both boundaries exhibit scale-free statistics (Fig.~\ref{fig:sup_events_characterization}.A) consistent with previous laboratory earthquake experiments~\cite{Lherminier2014,Lherminier2019}. As expected, the energy variations measured at the driving boundary are systematically larger than those recorded at the resisting boundary because friction along the confining cylinders dissipates part of the transmitted stress, leading simultaneously to larger apparent stiffnesses and torque variations near the driving plate (Fig.~\ref{fig:sup_events_characterization}.C,D).

\subsubsection*{Detection of Mechanical Avalanches}

Mechanical avalanches are identified from abrupt variations in the torque transmitted through the granular fault. To quantify the associated changes in elastic energy (see next section), both the driving and resisting torque signals are squared and analyzed through

\begin{equation}
\Delta\Gamma^{2}=\Gamma_f^{2}-\Gamma_i^{2},
\end{equation}

where $\Gamma_i$ and $\Gamma_f$ denote the torque immediately before and after the event. Simultaneously, the vertical motion of the upper plate is characterized by

\begin{equation}
\Delta h=h_f-h_i,
\end{equation}

with $\Delta h>0$ corresponding to dilation and $\Delta h<0$ to compaction.

Because a grain-scale rearrangement simultaneously modifies the stress transmitted through the packing and the sample thickness, avalanche detection is initially performed on both the driving and resisting torques simultaneously. All events detected in either signal are then stored, after which the corresponding displacement event is associated with the matched torque discontinuity.  The resulting event catalog therefore provides, for each avalanche, both the mechanical torque variation ($\Delta\Gamma^2$) and the associated volumetric response ($\Delta h$). The typical delay between synchronized signals is below 1~ms, whereas successive avalanches are separated by more than 100~ms, ensuring an unambiguous one-to-one correspondence between measurements. This common event catalog is subsequently used to match acoustic emissions and force-chain reorganizations.

Prior to event detection, the torque signals are first spectrally smoothed using a low-pass Fourier filter (\textsc{Tsmooth}, smoothing fraction $=0.1$), which suppresses high-frequency noise while preserving the characteristic loading dynamics. A second temporal smoothing, based on convolution with an adjustable window, controls the detection scale: narrow windows (1~ms) resolve the smallest avalanches, whereas broader windows (1~s) isolate only large-scale reorganizations.

Candidate events are then identified through a two-stage derivative analysis. Forward and backward numerical derivatives locate the onset and termination of each abrupt variation, while local refinement around each candidate determines the corresponding maxima and minima with sub-sample precision. Finally, consistency criteria enforce alternating start--end indices, chronological ordering, and one-to-one pairing between extrema, eliminating spurious detections. This procedure provides robust estimates of avalanche onset, duration, and amplitude over more than four decades of event size.

\subsubsection*{Evaluation of the Effective Stiffness}

Between two successive avalanches, the granular fault undergoes quasi-elastic loading under a constant driving angular velocity $\omega$. During these loading phases, the resisting torque increases approximately linearly with time,

\begin{equation}
\Gamma(t)=a\,t+b,
\end{equation}

where the slope

\begin{equation}
a=\frac{\mathrm{d}\Gamma}{\mathrm{d}t}
=R\omega k_{eff}
\end{equation}

is directly proportional to the effective stiffness $k$ of the loaded granular structure. Consequently, the stiffness is obtained from the local loading rate according to

\begin{equation}
k_{eff}=\frac{1}{R\omega}
\frac{\mathrm{d}\Gamma}{\mathrm{d}t}.
\label{eq:k_general}
\end{equation}

To estimate the local stiffness associated with each avalanche, all samples belonging to detected events are first removed from the torque signal. The remaining contiguous loading segments are identified using a cluster-detection algorithm and individually fitted by affine regression. For each loading interval, we extract the local slope $a$, intercept $b$, and the residual curvature, which quantifies departures from purely elastic loading. The values of the torque immediately before and after each avalanche are simultaneously stored for subsequent statistical analyses.

Applying Eq.~(\ref{eq:k_general}) to the different force measurements yields

\begin{equation}
k_{\mathrm{drive}}
=
\frac{1}{\omega R}
\frac{\sqrt{r^{2}+D^{2}}}{D}
a_{\mathrm{motor}},
\qquad
k_{\mathrm{grains}}
=
\frac{a_{\mathrm{resistive}}}
{\omega R},
\qquad
\end{equation}

where the geometric prefactor accounts for the chain transmission between the motor and the lower ring.

Because the apparatus (motor, transmission, and force sensor) is significantly stiffer than the granular layer, the measured compliance is dominated by the grains. Moreover, grain displacement fields (Fig.~\ref{fig:sup_img_analysis}) reveal that deformation localizes within two mechanically distinct regions: an actively sheared band adjacent to the driving boundary and a largely jammed region beneath the upper plate. The corresponding effective stiffnesses are therefore approximated by

\begin{equation}
k_{\mathrm{shear}}
\simeq
k_{\mathrm{drive}},
\label{eq:k_shear}
\end{equation}

and

\begin{equation}
k_{\mathrm{jammed}}
\simeq
k_{\mathrm{grains}},
\label{eq:k_jammed}
\end{equation}

respectively. Tracking the evolution of these local stiffnesses throughout the experiment provides a direct measure of the mechanical state of both the sheared and jammed regions before and after each avalanche.

\subsubsection*{Detection of Acoustic Bursts}

Acoustic emissions generated by grain-scale rearrangements are analyzed in the time--frequency domain using a Short-Time Fourier Transform (STFT) (\textit{18}, \textit{19}). The STFT is computed on overlapping windows of $2^{20}$ samples using a Hann taper to minimize spectral leakage. The resulting spectrogram is logarithmically transformed and smoothed with a Gaussian kernel spanning 12 frequency bins and 0.6~ms, enhancing transient acoustic activity while preserving temporal resolution.

For each time step, the spectral power is integrated over all frequencies to obtain a scalar measure of acoustic activity. Individual acoustic bursts are identified by thresholding this integrated signal, and their energy is computed as

\begin{equation}
E_a=\sum_{t\in ev} STFT(t),
\label{eq:Ea_def}
\end{equation}

where the sum extends over the duration of the detected event. Event amplitudes are additionally characterized by the peak value $\max\!\left[STFT(t\in ev)\right]$, providing a threshold-independent estimate of relative burst intensity.

To compensate for variations in sensor coupling and sensitivity, the acoustic energy recorded by each transducer is normalized by the total energy measured over the entire experiment and subsequently rescaled by the mean total energy across all sensors. Following this normalization, the energy distributions obtained from the six piezoelectric sensors collapse onto a common scaling form, demonstrating that the measured acoustic statistics are independent of sensor position and coupling. Minor deviations remain for the smallest events, consistent with differences in signal-to-noise ratio near the detection threshold.

Consistent with previous studies (\textit{2}), the acoustic energy distribution follows scale-free statistics over several decades (Fig.~\ref{fig:sup_events_characterization}.A). Mechanical and acoustic event populations exhibit nearly identical scaling behavior. To compare both observables at the event level, each mechanically detected avalanche is associated with the strongest acoustic burst occurring within a $\pm0.5$~s time window. The acoustic energy assigned to a given mechanical event is then defined as the maximum energy recorded among the six sensors, providing a robust estimate of the energy radiated during the corresponding granular reorganization.
 
\subsubsection*{Structural Response}

A photoelasticimetry setup is used to simultaneously resolve the grain-scale stress distribution and particle kinematics. The setup comprises a circular polarizer placed inside the inner acrylic cylinder, the birefringent grains, a second circular polarizer, and 24 cameras. This configuration provides simultaneous measurements of grain positions and intergranular stresses throughout the experiment.

The image-processing pipeline first reconstructs a panoramic view of the granular layer from the 24 individual camera images. Image registration is performed using the Scale-Invariant Feature Transform (SIFT) to detect and match common features between overlapping images. A reference panorama is manually assembled from a representative frame, and the resulting correspondences are used to determine a transformation matrix for each camera. These matrices remain fixed throughout the experiment and are subsequently used to project all images onto a common reference frame.

Grain detection and photoelastic stress measurements are performed on the individual camera images prior to projection. Grain centers are first identified using a two-step procedure based on the Hough transform. Their positions are then refined by pattern matching, maximizing the correlation between the grain image and a circular gradient template with the expected grain radius. This procedure provides robust localization and accurate grain-size estimation for both particle populations.

Potential grain contacts are identified using a geometrical criterion deliberately chosen to favor false positives over false negatives. Spurious contacts are subsequently rejected from the photoelastic signal based on the intensity variation along the putative contact. For each validated contact, the tangential force is estimated from the local photoelastic response using the contact luminosity and, when photoelastic fringes are resolved, the squared-gradient method~\cite{Daniels2017}.

Grain tracking between frames acquired before and after an event is facilitated by the limited displacement of grains in the compressed granular layer, which remains smaller than approximately one grain radius. Grain displacements are therefore determined by a least-squares matching procedure, providing a unique identifier for each particle across frames. Grains that cannot be matched are assigned new identifiers.

\subsubsection*{Force-Photoelasticity Calibration}

\begin{figure}
	\centering
	\includegraphics[width=6.5 in]{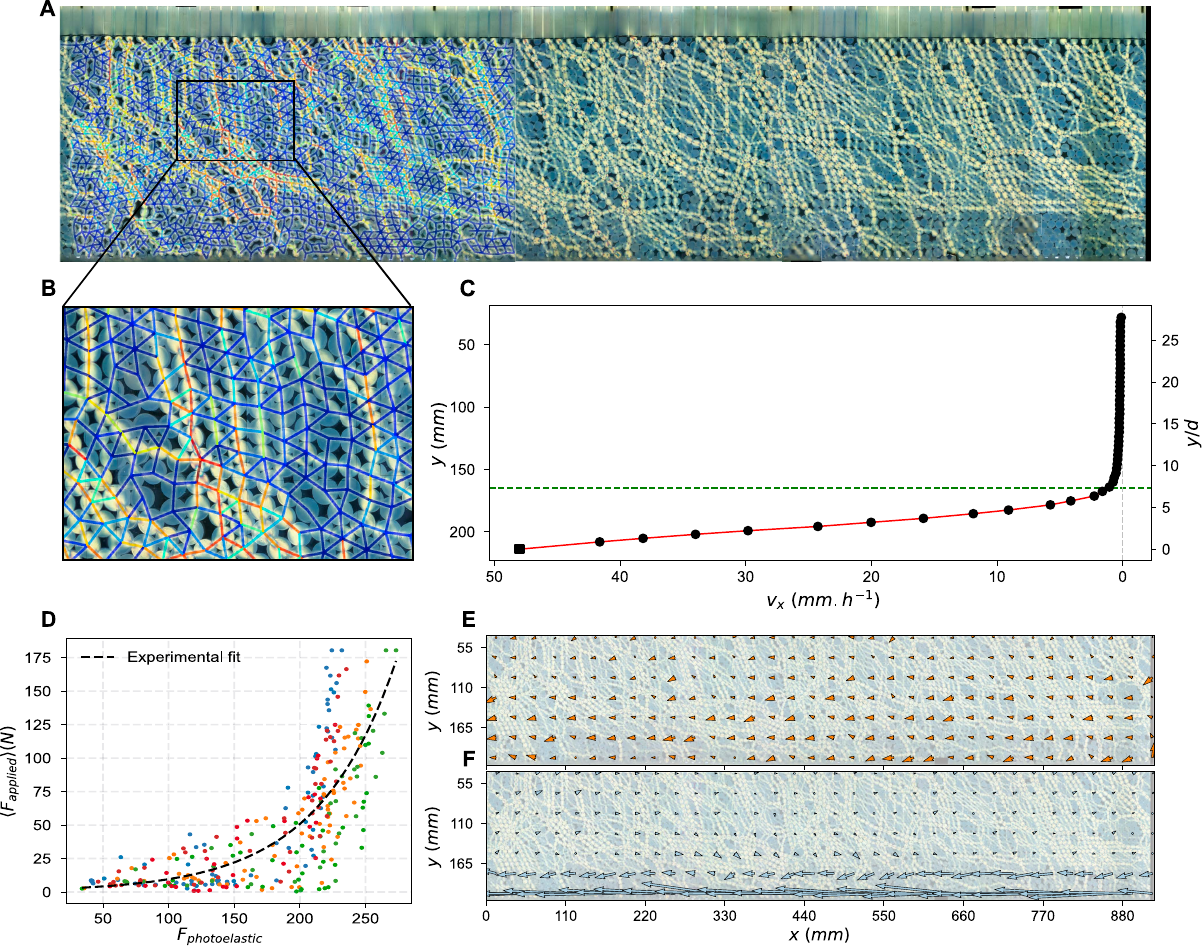} 
	\caption{\textbf{Shear-band to jammed zone transition}
		(\textbf{A}) Panoramic image reconstructed from the 24 camera images. In the left part, a portion of the contact network is superimposed on the image. (\textbf{B}) A zoom on a part of (\textbf{A}) to help visualize the estimated contact forces. (\textbf{C}) Velocity profile in the granular layer averaged over 35000 images, far from the transitory regime at the beginning of the experiment. The green dashed line represents the threshold between the jammed zone and the shear band. This value was fixed at around 8 diameters from the shearing wall: below this, the grains can flow; above is the jammed zone. (\textbf{D}) Experimental calibration of the photelastic response of the grains. 5 calibration grains have been used. All the data collapse on the experimental fit, represented by the dashed line. (\textbf{E}) (\textbf{F}) Coarse-grained displacement field average on 150 events for $Q2$ (light blue) and $Q3$ (orange), displayed on top of a reference panorama image. The x-axis is shared by both fields.}
	\label{fig:sup_img_analysis} 
\end{figure}

Photoelastic measurements are calibrated in situ using a dedicated setup. A pair of pliers equipped with a force sensor on the handle is inserted into the granular layer, and a single grain is placed between the jaws. The applied force is progressively increased while the force measured by the sensor is recorded. An image of the grain is acquired every second, allowing the photoelastic response to be directly related to the applied force (Fig.~\ref{fig:sup_img_analysis}.B). The resulting calibration curve is monotonic and is well described by
\begin{equation}
F_{\mathrm{calibrated}}
=
A\exp\left(bF_{\mathrm{photoelastic}}\right),
\end{equation}
with $A=1.80$ and $b=0.0167$. The same analysis pipeline is used for calibration and experimental measurements to ensure consistency between the two.

To quantify structural changes associated with major events, we compare the force-chain network immediately before and after each event. For each contact, the change in photoelastic force is calculated as
\begin{equation}
\Delta F_{\mathrm{photoelastic}}
=
F_{\mathrm{photoelastic}}(t_{\mathrm{after}})
-
F_{\mathrm{photoelastic}}(t_{\mathrm{before}}).
\end{equation}

The mean variation in contact strength over an event is then obtained by averaging over all $N_{\mathrm{contacts}}$ identified contacts:
\begin{equation}
\left\langle
\Delta F_{\mathrm{photoelastic}}
\right\rangle
=
\frac{1}{N_{\mathrm{contacts}}}
\sum_{\mathrm{contacts}}
\Delta F_{\mathrm{photoelastic}}.
\end{equation}









\begin{figure}
	\centering
	\includegraphics[width=6.5 in]{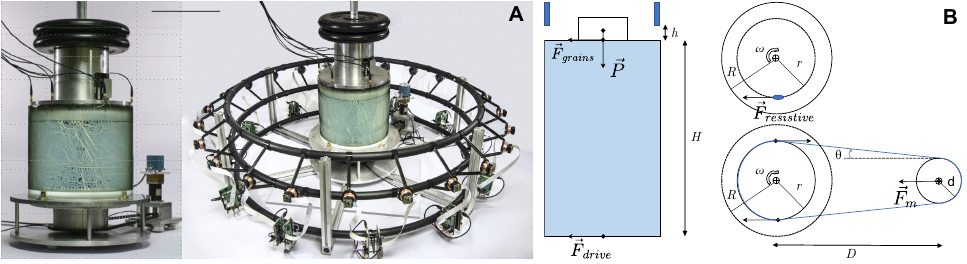} 
	\caption{\textbf{System apparatus}
		(\textbf{A}) Experimental apparatus consisting of a concentric acrylic cylinders containing oner layer of photoelastic grains, confined between a fixed upper plate and a motor-driven lower plate, which imposes the displacement. The lower plate is driven through a motor–chain mechanism, while the upper plate is instrumented with six acoustic sensors and two position sensors. Force sensors monitor both the resisting force at the upper plate and the driving force required to maintain the imposed displacement of the lower plate. A dead load is applied to the upper plate to confine the granular layer. Twenty-four cameras continuously image the granular assembly to resolve its structural evolution during shear. (\textbf{B}) Schematic of the experimental apparatus and the geometrical transformations used to convert the monitored forces into the relevant driving and granular forces, $\vec{F}{\mathrm{drive}}$ and $\vec{F}{\mathrm{grains}}$, respectively.}
	\label{fig:sup_experimental_set_up} 
\end{figure}

\end{document}